\documentclass[a4paper,reqno,final]{amsart}
\usepackage{amsmath,amssymb,amscd,eucal}
\usepackage[latin1]{inputenc}
\usepackage{array}
\usepackage{epsfig}
\usepackage{mathrsfs}
\usepackage{geometry}
\usepackage{hyperref}
\hypersetup{%
  pdftitle   = {Gauging the Wess--Zumino term of a sigma model with
    boundary},
  pdfkeywords = {Sigma model, gauge theory, boundary, equivariant
    cohomology, boundary WZW model, relative cohomology, generalised
    geometry, Courant bracket},
  pdfauthor  = {José Figueroa-O'Farrill, Noureddine Mohammedi},
  pdfcreator = {\LaTeX\ with package \flqq hyperref\frqq}
}
\newcommand{\into}{\hookrightarrow}
\newcommand{\eA}{\mathscr{A}}
\newcommand{\eL}{\mathscr{L}}
\newcommand{\RR}{\mathbb R}
\newcommand{\ZZ}{\mathbb Z}

\newcommand{\fg}{\mathfrak{g}}
\newcommand{\fh}{\mathfrak{h}}
\newcommand{\fk}{\mathfrak{k}}
\newcommand{\fn}{\mathfrak{n}}
\newcommand{\fsl}{\mathfrak{sl}}

\newcommand{\fS}{\mathfrak{S}}
\newcommand{\sH}{\mathscr{H}}
\newcommand{\sB}{\mathscr{B}}
\newcommand{\half}{\tfrac12}
\DeclareMathOperator{\Ad}{Ad}
\DeclareMathOperator{\id}{id}
\DeclareMathOperator{\Tr}{Tr}
\DeclareMathOperator{\dvol}{dvol}
\begin{document}

\title[Gauging Wess--Zumino terms with boundary]{Gauging the
  Wess--Zumino term of a sigma model with boundary}
\author[Figueroa-O'Farrill]{José Figueroa-O'Farrill}
\address{School of Mathematics, The University of Edinburgh, Scotland,
United Kingdom}
\email{j.m.figueroa@ed.ac.uk}
\author[Mohammedi]{Noureddine Mohammedi}
\address{Laboratoire de Mathématiques et Physique Théorique,
  Université de Tours, France}
\email{nouri@phys.univ-tours.fr}
\begin{abstract}
  We investigate the gauging of the Wess--Zumino term of a sigma model
  with boundary.  We derive a set of obstructions to gauging and we
  interpret them as the conditions for the Wess--Zumino term to extend
  to a closed form in a suitable equivariant relative de Rham complex.
  We illustrate this with the two-dimensional sigma model and we show
  that the new obstructions due to the boundary can be interpreted in
  terms of Courant algebroids.  We specialise to the case of the
  Wess--Zumino--Witten model, where it is proved that there always
  exist suitable boundary conditions which allow gauging any subgroup
  which can be gauged in the absence of a boundary.  We illustrate
  this with two natural classes of gaugings: (twisted) diagonal
  subgroups with boundary conditions given by (twisted) conjugacy
  classes, and chiral isotropic subgroups with boundary conditions
  given by cosets.
\end{abstract}
\thanks{EMPG-05-10}
\maketitle
\tableofcontents

\section{Introduction}

Ever since the advent of D-branes, field theories on manifolds with
boundary have attracted a great deal of attention.  One such theory,
with fascinating connections to geometry and topology, is the
nonlinear sigma model and in particular, in the context of string
theory, the two-dimensional nonlinear sigma model.  The sigma model
describes harmonic maps between two (pseudo)riemannian manifolds,
which we will call the spacetime and the target space.  Isometries of
the target space become classical global symmetries of the sigma model
and as with all classical symmetries we can try to gauge them, that
is, promote them to local symmetries by coupling the model to gauge
fields.  For a standard sigma model, whether or not the spacetime has
boundary, there is no obstruction in doing so; however things are
different when there is a (topologically nontrivial) Wess--Zumino (WZ)
term.

The gauging of the WZ term of a nonlinear sigma model was the subject
of much work in the late 1980s and early 1990s, spurred on by the
pivotal role played by the Wess--Zumino--Witten (WZW) model \cite{WW}
in a number of areas: string theory, conformal field theory and
integrable systems.  The two-dimensional case was investigated
independently in \cite{HS2d,JJMO} and later generalised to higher
dimension in \cite{HS}.  As the dimension grows the number of
obstructions increases and a very clear pattern emerges.  This was
interpreted in terms of equivariant cohomology independently in
\cite{WuEquivariant,FSsigma,FSec}; although see also \cite{W} for the
gauging of the WZW model.

The purpose of this note is to re-examine this problem when the
spacetime has boundary.  The WZ term of a sigma model with
boundary is a straight-forward generalisation of the case of the
boundary WZW model treated in \cite{KlS,Gaw,FSrc} and will be reviewed
below.  We extend the interpretation of gauging the WZ term in terms
of a relative version of equivariant cohomology and in this way derive
a number of new obstructions.  Remarkably, for the case of the WZW
model, these new obstructions are automatically overcome and we will
show that we can gauge \emph{any} subgroup which can be gauged in the
absence of a boundary, provided that the boundary conditions are
chosen appropriately.  This extends some recent results
\cite{GawedzkiCoset,ElitzurSarkissian,KRWZ} on gauged WZW models with
boundary conditions given by (twisted) conjugacy classes.  For the
case of the two-dimensional sigma model we interpret the new
obstructions in terms of the existence of subalgebroids of the Courant
algebroid of the target and of its restriction to the boundary
submanifold.

This note is organised as follows.  In Section~\ref{sec:relative} we
recall the definition of the WZ term in a sigma model with boundary.
We recall the topological conditions needed for the physical
consistency of such a term.  In Section~\ref{sec:equivariant} we
recall the obstructions to gauging a WZ term in the case of a sigma
model without boundary and their interpretation in terms of
equivariant cohomology and in particular their derivation using the
Cartan model for the equivariant cohomology.  In
Section~\ref{sec:both} we derive the obstructions to gauging the WZ
term in a sigma model with boundary.  This is derived using the Cartan
model for what could be termed relative equivariant cohomology.  We
are aware of two competing complexes which could claim to compute the
relative equivariant cohomology and we explain this situation in
Appendix~\ref{app:equivrel}.  We illustrate our results throughout
with the case of a two-dimensional sigma model.  We also re-interpret
the conditions for gauging in terms of Courant algebroids, extending
this observation to arbitrary dimension and to the presence of
boundary.  Finally, in Section~\ref{sec:wzw} we apply these
considerations to the WZW model and show that we may gauge \emph{any}
subgroup which can be gauged in the absence of boundary, provided we
choose compatible boundary conditions, namely orbits of the action of
the group we are trying to gauge.  We illustrate this with two natural
classes of gaugings: twisted diagonal gaugings with boundary
conditions given by twisted conjugacy classes, and chiral isotropic
gaugings with boundary conditions given by cosets.

\section{The WZ term of a sigma model with boundary}
\label{sec:relative}

In this section we review the definition of the WZ term for
a sigma-model with boundary and the conditions of the well-definedness
of the path integral.  We follow the treatment in \cite{FSrc}, to
where the reader is referred for more details.

Let $\Sigma$ be an oriented riemannian $d$-dimensional manifold and
$X$ be an oriented pseudo-riemannian manifold with $\dim X = n$.  Let
us first assume that $\Sigma$ has no boundary.  The sigma model is a
theory of maps $\varphi:\Sigma \to X$.  Let $g$ denote the metric of
$X$, $\eta$ the metric on $\Sigma$ and $\star$ the Hodge star
operation on $\Sigma$.  Relative to a local coordinate chart $x^i$ on
$X$ and letting $\varphi^i = x^i \circ \varphi$ denote the component
functions, the sigma model action takes the form
\begin{equation}
  \label{eq:sigma}
  S_\sigma[\varphi] := \tfrac12 \int_\Sigma g_{ij}(\varphi) d\varphi^i
  \wedge \star d\varphi^j = \tfrac12 \int_\Sigma g_{ij}(\varphi)
  \partial_\alpha \varphi^i \partial_\beta \varphi^j
  \eta^{\alpha\beta} \dvol_\Sigma~.
\end{equation}
In more invariant terms, it is (one half of) the $L^2$ norm of
$d\varphi \in \Omega^1(\Sigma,\varphi^*TX)$.

To this action it might be possible to add a WZ term.  Let
$H$ be a closed $(d+1)$-form on $X$.  Let $\varphi(\Sigma) \subset X$
denote the image of $\Sigma$ under the map $\varphi$.  Since $\Sigma$
has no boundary, neither does its image.  If there is a
$(d+1)$-dimensional submanifold (or indeed a $(d+1)$-chain) $M \subset
X$ whose boundary is $\varphi(\Sigma)$ then we can add a WZ
term to the action
\begin{equation}
  \label{eq:sigmawz}
  S_{\text{WZ}} := \int_M H~.
\end{equation}
Since $dH=0$ the variation of the WZ term is a boundary term
\begin{equation*}
  \delta S_{\text{WZ}} = \int_M \eL_{\delta\varphi} H = \int_M d
  \imath_{\delta \varphi} H = \int_{\varphi(\Sigma)}
  \imath_{\delta\varphi} H~,
\end{equation*}
whence its contribution to the equations of motion involve only the
original map $\varphi:\Sigma \to X$.

The existence of $M$ requires that the cycle $\varphi(\Sigma)$ bounds,
whence its homology class $[\varphi(\Sigma)] \in H_d(X)$ vanishes.
This has to be true for all maps $\varphi$ and this is guaranteed by
$H_d(X) = 0$.  The archetypal example is the WZW
model when $d=2$ and $X$ is a simply-connected compact simple Lie
group, $g$ a bi-invariant metric and $H$ the standard bi-invariant
three-form.  It is a theorem of Cartan that $H_2(X) = 0$ in this
case.

There is also the question of the dependence on $M$.  As seen above,
the classical equations of motion are independent of $M$, but the
quantum theory in principle does depend on the choice of $M$.  Indeed,
suppose that $M' \subset X$ is another chain whose boundary is
$\varphi(\Sigma)$.  The difference between the WZ terms
corresponding to $M$ and $M'$ is
\begin{equation*}
  \Delta S_{\text{WZ}} = \int_M H - \int_{M'} H = \int_N H~,
\end{equation*}
where $N$ is the cycle obtained by gluing $M$ and $M'$ (with the
opposite orientation) along their common boundary $\varphi(\Sigma)$.
Thus $N$ defines a class $[N] \in H_{d+1}(X)$ and $\Delta
S_{\text{WZ}}$ is the evaluation of the (de Rham) cohomology class $[H]
\in H^{d+1}(X)$ on $[N]$.  The path integral will be independent of
the choice of $M$ provided that $\Delta S_{\text{WZ}} \in 2 \pi \ZZ$,
or equivalently, when $[H/2\pi]$ defines an integral cohomology
class.

To summarise, there is an obstruction in $H_d(X)$ to defining the WZ
term and a quantisation condition $\frac1{2\pi}[H] \in H^{d+1}(X;\ZZ)$
to ensure that the (quantum) theory is well defined.

Now let us suppose that $\Sigma$ has boundary $\partial\Sigma$.  We
specify boundary conditions by demanding that $\varphi$ should map
$\partial\Sigma$ to a given submanifold $Y \subset X$.  We will
abbreviate this by saying that $\varphi:(\Sigma,\partial\Sigma) \to
(X,Y)$, suggesting a ``relative'' formulation of the problem.  We can
still write the sigma model term $S_\sigma$, but we immediately come
to a problem with the WZ term: since $\Sigma$ has boundary,
so will its image $\varphi(\Sigma)$ and hence there will be no $M$
with $\partial M = \varphi(\Sigma)$.  The way out is to relativise the
problem modulo $Y$; that is, we only demand that $\partial M =
\varphi(\Sigma)$ \emph{modulo} $Y$.  In other words, we demand that
$\partial M = \varphi(\Sigma) + D$ with some chain $D \subset Y$.  The
obstruction to the existence of such $(M,D) \subset (X,Y)$ is the
relative homology class $[\varphi(\Sigma)] \in H_d(X,Y)$.  (See
Appendix~\ref{app:equivrel} for a brief scholium on relative
(co)homology.)

Now suppose that there exists a $d$-form $B$ on $Y$ such that $i^*H =
dB$, where $i:Y \into X$ denotes the embedding.  This means that
$(H,B)$ defines a relative (de Rham) cohomology class $[(H,B)] \in
H^{d+1}(X,Y)$.  We can then write the following ``relative''
WZ term
\begin{equation}
  \label{eq:wzbdry}
  S_{\text{rWZ}} = \int_M H - \int_D B~.
\end{equation}
The $d$-form $B$ does not enter in the equations of motion, but
only in specifying the boundary conditions.  Indeed, the variation of
the relative WZ term is
\begin{equation*}
  \begin{split}
    \delta S_{\text{rWZ}} & = \int_M \eL_{\delta\varphi} H - \int_D
    \eL_{\delta\varphi} B\\
    & = \int_{\partial M} \imath_{\delta\varphi} H  - \int_D
    \imath_{\delta\varphi} d B - \int_{\partial D}
    \imath_{\delta\varphi} B\\
    & = \int_{\varphi(\Sigma)} \imath_{\delta\varphi} H +
    \int_{\varphi(\partial\Sigma)} \imath_{\delta\varphi}B~,
  \end{split}
\end{equation*}
where we have used that on $D$, $i^*H = dB$ and that $\partial D
= - \partial \varphi(\Sigma) = - \varphi(\partial\Sigma)$.

There is also a quantisation condition which guarantees the
independence of the path integral on the choice of $(M,D)$.  Suppose
that $(M',D') \subset (X,Y)$ is such that $\partial M' =
\varphi(\Sigma) + D'$.  The difference in the WZ terms is now
\begin{equation*}
  \Delta S_{\text{rWZ}} = \int_M H - \int_{M'} H  - \int_D B +
  \int_{D'} B = \int_N H - \int_{\partial N} B~,
\end{equation*}
where $N = M-M'$ is the chain obtained by gluing $M$ and $M'$ (with
the opposite orientation) along the common part of their boundary
$\varphi(\Sigma)$.  This still leaves some boundary $\partial N$ which
is easily calculated to be $D - D'$.  This means that $(N,\partial N)$
defines a relative homology class $[(N,\partial N)] \in
H_{d+1}(X,Y)$.  The change in the WZ is then the evaluation
of the relative (de Rham) cohomology class $[(H,B)]$ on
$[(N,\partial N)]$ and the path integral will not see this provided
that the result is in $2\pi\ZZ$, which implies the integrality of the
relative cohomology class $[(H,B)]/2\pi$.

To summarise, there is an obstruction in $H_d(X,Y)$ to defining the WZ
term and a quantisation condition $\tfrac1{2\pi}[(H,B)] \in
H^{d+1}(X,Y;\ZZ)$ to ensure that the theory is well defined.  In other
words, the situation is as in the case without boundary but now
relative to $Y$.

\section{Gauging the WZ term}
\label{sec:equivariant}

In this section we review the gauging of the WZ term in a
sigma model without boundary and its relation with equivariant
cohomology.  We follow the treatment in \cite{FSec,FSsigma}.

We start with the setup up of the previous section, where $\Sigma$ has
no boundary.  We assume throughout that $G$ is a connected Lie group
acting on $X$ in such a way that both the metric $g$ and $H$ are
preserved, so that the action $S_\sigma + S_{\text{WZ}}$ is
$G$-invariant.  We can then set out to gauge the symmetry.  Let $\fg$
denote the Lie algebra of $G$.  To every $U\in \fg$ there corresponds
a Killing vector $\xi_U$ on $X$ which in addition satisfies that
$d\imath_{\xi_U} H = 0$, since $H$ is both closed and invariant.  The
sigma model term $S_\sigma$ can be gauged simply by minimal coupling
to a gauge field $A$; that is, to a family of locally defined
$\fg$-valued one-forms on $\Sigma$.  Minimal coupling consists in
substituting the exterior derivative for a covariant exterior
derivative:
\begin{equation}
  \label{eq:minimal}
  d\varphi^i \mapsto d^\nabla\varphi^i := d\varphi^i - A^a
  \xi_a^i(\varphi)~,
\end{equation}
where $A^a$ are the components of $A$ relative to a basis $U_a$ for
$\fg$ and $\xi_a^i$ is the $i$-th component of the Killing vector
$\xi_{U_a}$.  The gauged sigma model term is then
\begin{equation}
  \label{eq:sigmagauged}
  S_{\text{g}}[\varphi,A] := \tfrac12 \int_\Sigma g_{ij}(\varphi)
  d^\nabla\varphi^i \wedge \star d^\nabla\varphi^j = \tfrac12
  \int_\Sigma g_{ij}(\varphi) \nabla_\alpha\varphi^i
  \nabla_\beta\varphi^j \eta^{\alpha\beta} \dvol_\Sigma~,
\end{equation}
where we have written $\nabla_\alpha\varphi^i = \partial_\alpha
\varphi^i - A^a_\alpha \xi_a^i$.

Gauging the WZ term is a different matter: minimal coupling
$H$ does not generally result in a theory with local equations of
motion, since the resulting form is not in general closed.  Indeed,
gauging the WZ term is hindered by a set of obstructions
\cite{HS2d,JJMO,HS} which can be described succinctly in terms of
equivariant cohomology \cite{FSsigma,FSec}.  The statement is that the
WZ term can be gauged if and only if $H$ extends to an
equivariant closed form.  Let us explain this statement.

Under an infinitesimal gauge transformation with parameter $\lambda:
\Sigma \to \fg$, a differential form $H \in \Omega(X)$ transforms as
follows
\begin{equation}
  \label{eq:infgautra}
  \delta_\lambda H = \lambda^a \eL_a H + d\lambda^a \wedge \imath_a
  H~,
\end{equation}
where $\imath_a$ and $\eL_a$ are the contraction of and Lie derivative
along the Killing vector $\xi_{U_a}$, respectively, and $\lambda^a :
\Sigma \to \RR$ are the component functions of $\lambda$ with respect
to the basis $U_a$.  Gauging means coupling to a gauge field $A$,
which is locally a one-form on $\Sigma$ with values in $\fg$, and
which transforms under an infinitesimal gauge transformation as
\begin{equation*}
  \delta_\lambda A^a= d\lambda^a - f_{bc}{}^a \lambda^b A^c~.
\end{equation*}
The curvature two-form $F = dA + \half [A,A]$ transforms as
\begin{equation*}
  \delta_\lambda F^a = - f_{bc}{}^a \lambda^b F^c~.
\end{equation*}
It is possible to write these transformations in a way analogous to
\eqref{eq:infgautra} by introducing the contractions
\begin{equation}
  \imath_a A^b = \delta_a{}^b \qquad\text{and}\qquad \imath_a F^b =
  0
\end{equation}
and the Lie derivatives $\eL_a = d \imath_a + \imath_a d$.  In this
way we have that the infinitesimal gauge transformation of \emph{any}
expression involving $A$ and $F$ and differential forms on $X$ is
given by \eqref{eq:infgautra} from where one reads that such an
expression is gauge-invariant provided that it is annihilated by both
$\imath_a$ and $\eL_a$ for all $a$.  We can formalise this as follows.

Let $\Omega^\bullet(X)$ denote the de Rham complex of differential
forms on $X$.  It is a differential graded algebra (dga) relative to
the exterior derivative and the wedge product.  The $G$-action on $X$
induces one on $\Omega^\bullet(X)$ and because $G$ is connected, this
action is trivial on cohomology.  At the level of the Lie algebra,
this can be seen as follows.  Every element $U_a$ of the Lie algebra
defines an antiderivation $\imath_a$ corresponding to contraction with
the Killing vector $\xi_{U_a}$, and also defines a derivation $\eL_a =
[d,\imath_a]$, which shows that $\eL_a$ acts trivially in the
cohomology.  One has the usual formulae $[\eL_a,\imath_b] = f_{ab}{}^c
\imath_c$ and $[\eL_a,\eL_b] = f_{ab}{}^c \eL_c$.  The existence of
these (anti)derivations turns $\Omega^\bullet(X)$ into a $G$-dga.  Now
let $(\eA,d,\imath)$ be \emph{any} $G$-dga.  We say that an element
$\phi\in \eA$ is horizontal if $\imath_a \phi = 0$ for all $a$.
Similarly we say that $\phi \in \eA$ is invariant if $\eL_a \phi = 0$
for all $a$.  If $\phi$ is both horizontal and invariant we say that
it is basic.  Note that if $\phi$ is both horizontal and closed
($d\phi =0$) then it is automatically invariant and hence basic.

In any $G$-dga, the basic elements form a subcomplex and hence a
differential graded subalgebra.  The archetypal example of a $G$-dga
is the Weil algebra $W(\fg)$ which is freely generated by
an abstract $\fg$-valued one-form $A$ and an abstract $\fg$-valued
two-form $F$ subject to the following relations
\begin{equation}
  \label{eq:Weil}
  dA = -\tfrac12 [A,A] + F \qquad\text{and}\qquad dF = [F,A]
\end{equation}
modelled on the structure equations for a connection on a principal
$G$-bundle.  In fact, this is an algebraic model for the de Rham
complex of the total space of the universal bundle $EG \to BG$ over
the classifying space of the group $G$.  The contractibility of $EG$ is
reflected in the fact that $W(\fg)$ is acyclic.  The antiderivation
$\imath_a$ is defined by $\imath_a A = U_a$ and $\imath_a F = 0$, and
we define $\eL_a = [d,\imath_a]$.

We define $\Omega_G^\bullet(X)$ as the basic subcomplex of the tensor
product $W(\fg) \otimes \Omega(X)$, which is naturally a $G$-dga.  The
cohomology $H_G(X)$ is called the $G$-equivariant cohomology of
$M$.  As seen above, the basic forms in $W(\fg) \otimes \Omega(X)$ are
precisely the gauge-invariant terms in the corresponding sigma model.
The gauging of the WZ term consists in extending the
WZ term $H$ to a gauge-invariant term which in addition is
closed; that is, to an equivariant cocycle.  That is, we seek
\begin{equation}
  \label{eq:ext}
  \sH = H + \phi_a \wedge A^a + \theta_{a} \wedge F^a + \tfrac12
  \phi_{ab} \wedge A^a \wedge A^b + \cdots \in \Omega_G^{d+1}(X)~,
\end{equation}
where
\begin{equation*}
  \Omega_G^{d+1}(X) = \bigoplus_{p=0}^{d+1} W^p(\fg) \otimes
  \Omega^{d+1-p}(X)~,
\end{equation*}
such that $d\sH = 0$.

The simplest way to derive the explicit expressions for the
obstruction to gauging the WZ term is to work with the
Cartan model for equivariant cohomology.  The departing observation is
that the dependence on the gauge field $A$ of a local gauge invariant
expression is via the exterior covariant derivative (or $F$), whence
we should be able to dispense with $A$.  The way to do this is to
introduce the Cartan model for equivariant cohomology.  The basic
subcomplex of $W(\fg) \otimes \Omega(X)$ is isomorphic to the complex
whose cochains are the $G$-invariants in $\fS\fg^* \otimes \Omega(X)$
relative to a twisted differential $d_C$ defined by
\begin{equation*}
  d_C \phi = d\phi - F^a \imath_a \phi \qquad\text{and}\qquad d_C F =
  0~,
\end{equation*}
where $\phi\in\Omega(X)$.  The map
\begin{equation*}
  \left(W(\fg) \otimes \Omega(X)\right)_{\text{basic}}
  \xrightarrow{\cong} \left(\fS\fg^* \otimes \Omega(X)\right)^G
\end{equation*}
is given simply by putting $A=0$, whereas the inverse map is given by
minimal coupling.  Therefore we will write the Cartan representative
for $\sH$ as
\begin{equation*}
  \sH_C = H + \theta_a F^a + \half \theta_{ab} F^a F^b + \cdots
\end{equation*}
with $\eL_a \theta_{b} = f_{ab}{}^c \theta_c$, et cetera, and we
demand that $d_C \sH_C = 0$, which gives the sequence of conditions
\begin{equation*}
  \imath_a H = d\theta_a,\qquad
  \imath_a\theta_b + \imath_b \theta_a = d\theta_{ab},\qquad
  \imath_a\theta_{bc} + \text{cyclic} = d\theta_{abc},\qquad \ldots
\end{equation*}
In each equation the left-hand side is closed and the obstruction to
gauging is the obstruction of that closed form being exact.

For example, for a two-dimensional sigma model, which is the case we
are primarily interested in, the Cartan representative for the gauged
WZ term takes the form
\begin{equation*}
  \sH_C = H + \theta_a F^a~,
\end{equation*}
for some $1$-forms $\theta_a$ satisfying $\eL_a \theta_b = f_{ab}{}^c
\theta_c$.   The obstructions to gauging are given by
\begin{equation*}
  d_C \sH_C = - \imath_a H F^a +  d\theta_a F^a - \imath_a \theta_b
  F^a F^b = 0~,
\end{equation*}
or equivalently
\begin{equation}
  \label{eq:2dcocycle}
  \imath_a H = d\theta_a \qquad\text{and}\qquad
  \imath_a \theta_b = - \imath_b \theta_a~,
\end{equation}
which in components become
\begin{equation*}
  \xi^i_a H_{ijk} = \partial_j \theta_{a\,k} - \partial_k
  \theta_{a\,j} \qquad\text{and}\qquad
  \xi_a^i \theta_{b\, i} = -   \xi_b^i \theta_{a\, i}~.
\end{equation*}
Minimally coupling $\sH_C$ we obtain
\begin{equation*}
    \sH = H - \imath_a H A^a - \half \imath_a\imath_bH A^a \wedge A^b
    + \tfrac16\imath_a\imath_b\imath_cH A^a\wedge A^b \wedge A^c
    + \theta_a F^a - \imath_a \theta_b A^a \wedge F^b
\end{equation*}
which, using the relations \eqref{eq:2dcocycle}, can be simplified to
\begin{equation}
  \label{eq:2dGWZWterm}
    \sH = H + d\left( A^a \theta_a + \half \imath_a \theta_b A^a
      \wedge A^b \right)~.
\end{equation}
As a result, the gauged WZ term is given by
\begin{align*}
  S_{\text{gWZ}} &= \int_M H + \int_{\varphi(\Sigma)} \left( A^a
    \theta_a + \half \imath_a\theta_b A^a \wedge A^b
  \right)\\
  &= \int_{\widetilde\Sigma} \tfrac16 H_{ijk}(\varphi)
  \partial_\alpha\varphi^i \partial_\beta\varphi^j
  \partial_\gamma\varphi^k \varepsilon^{\alpha\beta\gamma} d^3\sigma
  \\
  & \qquad + \int_\Sigma \left( A_\alpha^a \theta_{a\, i}(\varphi)
    \partial_\beta\varphi^i + \half \xi_a^i \theta_{b\, i}(\varphi)
    A_\alpha^a A_\beta^b \right) \varepsilon^{\alpha\beta} d^2\sigma~,
\end{align*}
where $\widetilde\Sigma$ is a three-manifold with boundary $\Sigma$,
to which the $\varphi^i$ have been extended.

\section{Gauging the WZ term with boundary}
\label{sec:both}

In this section we present the obstructions to gauging the
WZ term in a sigma model with boundary and, for the two-dimensional
case, interpret them in terms of Courant brackets.

\subsection{The obstructions from the boundary}

As recalled above, the WZ term of a sigma model with
boundary $\varphi: (\Sigma, \partial\Sigma) \to (X,Y)$ is given by
\begin{equation*}
  \int_M H - \int_D B
\end{equation*}
where $D \subset Y$, $\partial M = \varphi(\Sigma) + D$, $i^* H =
dB$ and $i:Y \to X$ is the embedding.  Let $G$ act on $X$ in such
a way that it preserves the boundary conditions, that is, the
submanifold $Y$.  We will further assume that $G$ does not just
preserve $H$ but also $B$.  (This means that $i^*[H] = 0$ also in
the $G$-invariant cohomology $H^{d+1}(Y)^G$.)  Then we will have
gauged this term if we can extend $H$ to a closed equivariant form
$\sH$ in such a way that $i^* \sH = d \sB$, with $\sB$ an
extension of $B$ by terms depending on the gauge field.
Equivalently, in the Cartan model, $i^*\sH_C = d_C \sB_C$.  The
gauged WZ term is then
\begin{equation*}
  S_{\text{grWZ}} = \int_M \sH - \int_D \sB~.
\end{equation*}

As an illustration let us consider the two-dimensional case.  In the
Cartan model
\begin{equation*}
  \sH_C = H + \theta_a F^a \qquad\text{and}\qquad \sB_C = B +
  h_a F^a~,
\end{equation*}
for some functions $h_a$ and one-forms $\theta_a$ satisfying $\eL_a
h_b = f_{ab}{}^c h_c$ and $\eL_a \theta_b = f_{ab}{}^c \theta_c$.  As
we saw above, $H$ extends to an equivariant closed form if the
conditions \eqref{eq:2dcocycle} are satisfied.  In addition, the
relative condition $i^*\sH_C = d_C \sB_C$ expands to
\begin{equation}
  \label{eq:2drelative}
  i^* H = d B \qquad\text{and}\qquad i^*\theta_a = dh_a -
  \imath_a B~.
\end{equation}
Notice that $d(i^*\theta_a + \imath_a B) = 0$ because of invariance of
$B$, and the condition is that this closed form should be exact, that
is, $dh_a$.

Minimally coupling $\sB_C$ we obtain
\begin{equation*}
    \sB = B + \imath_a B A^a - \half
    \imath_a\imath_bB A^a\wedge A^b + h_a F^a~,
\end{equation*}
which, using \eqref{eq:2drelative}, can be simplified to
\begin{equation}
    \sB = B + i^*\left( A^a \theta_a + \half \imath_a
      \theta_b A^a \wedge A^b \right) + d (h_a A^a)~.
\end{equation}
As a result, the gauged relative WZ term is given by
\begin{equation}
  \label{eq:GRWZ}
  \begin{split}
    S_{\text{grWZ}} &= \int_M H - \int_D B + \int_{\varphi(\Sigma)}
    \left( A^a \theta_a + \half \imath_a\theta_b A^a \wedge A^b \right)
    +
    \int_{\varphi(\partial\Sigma)} h_a A^a\\
    &= \int_{\widetilde\Sigma} \tfrac16 H_{ijk}(\varphi)
    \partial_\alpha\varphi^i \partial_\beta\varphi^j
    \partial_\gamma\varphi^k \varepsilon^{\alpha\beta\gamma} d^3\sigma\\
    & \qquad + \int_\Sigma \left( A_\alpha^a \theta_{a\, i}(\varphi)
      \partial_\beta\varphi^i + \half \xi_a^i \theta_{b\, i}(\varphi)
      A_\alpha^a A_\beta^b \right) \varepsilon^{\alpha\beta} d^2\sigma\\
    & \qquad - \int_\Delta \half B_{ij}(\varphi) \partial_\alpha
    \varphi^i \partial_\beta \varphi^j \varepsilon^{\alpha\beta}d^2\zeta
    - \int_{\partial\Sigma} h_a(\varphi) A^a_\sigma d\sigma~,
  \end{split}
\end{equation}
where now $\widetilde\Sigma$ is a 3-manifold with boundary
$\partial\widetilde\Sigma = \Sigma + \Delta$, and where we have
extended the maps $\varphi^i$ to $\widetilde\Sigma$.  The local
coordinates on $\Delta$ are called $\zeta$ and the local coordinate on
$\partial\Sigma$ is called $\sigma$.  We notice that, as in the case
without boundary, the gauge fields appear algebraically.

\subsection{Gauged WZ terms and Courant brackets}
\label{sec:courant}

As observed in \cite{AlekseevStrobl}, there is an interpretation of
the conditions \eqref{eq:2dcocycle} for gauging the WZ term of a
two-dimensional sigma model in terms of Courant brackets and the
generalised geometry of the target $X$, that is the geometry of $TX
\oplus T^*X$.  Indeed, if a symmetry group $G$ of the WZ term given by
$H \in \Omega^3(X)$ can be gauged, then the image of the map
\begin{equation*}
  \fg \to C^\infty(TX \oplus T^*X) \qquad\text{defined by}\qquad
  U_a \mapsto \xi_a + \theta_a
\end{equation*}
is isotropic and involutive under the $H$-twisted Courant bracket
(see, e.g., \cite[Section~3.7]{Gualtieri} and references therein)
\begin{equation}
  \label{eq:Courant}
  [v + \alpha, w + \beta]_H := [v,w] + \eL_v\beta - \eL_w\alpha -
  \half d(\imath_v \beta - \imath_w \alpha) - \imath_v\imath_w H~,
\end{equation}
for all $v+\alpha,w+\beta \in C^\infty(TX \oplus T^*X)$.  Indeed,
\begin{equation*}
  \begin{split}
    [\xi_a + \theta_a , \xi_b + \theta_b]_H &= [\xi_a,\xi_b] + \eL_a
    \theta_b - \eL_b \theta_a - \half d(\imath_a\theta_b - \imath_b
    \theta_a) - \imath_a\imath_b H \\
    &= f_{ab}{}^c \xi_c + \half f_{ab}{}^c \theta_c - \half f_{ba}{}^c
    \theta_c +  \half \imath_a d\theta_b - \half \imath_b d\theta_a -
    \imath_a d\theta_b\\
    &= f_{ab}{}^c (\xi_c + \theta_c)~,
  \end{split}
\end{equation*}
whereas isotropy is simply the condition $\imath_a\theta_b +
\imath_b\theta_a =0$.  In other words, there is Lie subalgebroid
(isomorphic to $\fg$) of the twisted Courant algebroid on $X$
associated to $H$.  Except for the maximality condition, we might call
this a twisted Dirac structure on $X$.

This result holds also in higher dimensions.  Instead of the Courant
bracket on $TX \oplus T^*X$ we have to consider its generalisation to
$TX \oplus \Lambda^{d-1}T^*X$, also called the Vinogradov bracket
(see, e.g., \cite{BoZa}).  The expression for this bracket is formally
identical to that in \eqref{eq:Courant} except that $\alpha,\beta \in
\Omega^{d-1}(X)$.  The proof of involutivity is identical to the case
of $d=2$ above.  The only point to notice is that $\imath_a \theta_b +
\imath_b \theta_a = d\theta_{ab}$, which suggests broadening of the
definition of the notion of isotropy, where $\imath_a\theta_b +
\imath_b\theta_a$ need not vanish, but merely be exact, which points
in the direction of $A_\infty$ structures.  Of course for $d=2$ there
is no such freedom since there are no exact $0$-forms.  This weaker
notion of isotropy still ensures the Jacobi identity for involutive
sub-bundles.  Hence gaugings give rise to Lie subalgebroids
(isomorphic to $\fg$) of the twisted Vinogradov algebroid on $X$
associated to $H$.

The new obstructions due to the presence of the boundary also have a
similar interpretation.  First of all, the condition that $i^*H = dB$,
says that $(Y,B) \subset (X,H)$ is a generalised submanifold in the
language of \cite[Definition~7.4]{Gualtieri} and
\cite[Definition~4]{BoZa}.  Now consider the map
\begin{equation*}
  \fg \to C^\infty(TY \oplus \Lambda^{d-1}T^*Y) \qquad \text{defined
    by} \qquad U_a \mapsto \xi_a + i^*\theta_a~,
\end{equation*}
where we have used that $\xi_a$ are tangent to $Y$.  Let's compute
their twisted bracket (on $Y$):
\begin{equation*}
  [\xi_a + i^*\theta_a, \xi_b+ i^*\theta_b]_{i^*H}~.
\end{equation*}
Since $i^*H = dB$, we have that
\begin{equation*}
  [\xi_a + i^*\theta_a, \xi_b+ i^*\theta_b]_{i^*H} = [e^B(\xi_a +
  i^*\theta_a), e^B(\xi_b+ i^*\theta_b)]~,
\end{equation*}
where the bracket on the right-hand side is the untwisted Courant
bracket (simply put $H=0$ in equation~\eqref{eq:Courant}) and where
the B-field transform $e^B : C^\infty(TY \oplus \Lambda^{d-1}T^*Y) \to
C^\infty(TY \oplus \Lambda^{d-1}T^*Y)$ is defined by
\begin{equation*}
  e^B(V + \alpha) = V + \alpha + \imath_V B~.
\end{equation*}
In our case,
\begin{equation*}
  e^B(\xi_a + i^*\theta_a) = \xi_a + dh_a~,
\end{equation*}
where we have used equation \eqref{eq:2drelative}, which holds also
for $d>2$.  Then we have
\begin{equation*}
  \begin{split}
    [e^B(\xi_a + i^*\theta_a), e^B(\xi_b+ i^*\theta_b)] &= [\xi_a +
    dh_a, \xi_b + dh_b]\\
    &= [\xi_a, \xi_b] + \eL_a dh_b - \eL_b dh_a - \half d (\imath_a
    dh_b - \imath_b dh_a)\\
    &= [\xi_a,\xi_b] + \eL_a dh_b \\
    &= f_{ab}{}^c (\xi_c + dh_c)\\
    &= f_{ab}{}^c e^B(\xi_c + i^* \theta_c)~,
  \end{split}
\end{equation*}
where we have used that $\eL_a d = d \eL_a$.  Therefore we have again
a Lie subalgebroid of the canonical Vinogradov algebroid on $Y$, which
is isomorphic to $\fg$ under the map
\begin{equation*}
  \fg \to C^\infty(TY \oplus \Lambda^{d-1}T^*Y) \qquad\text{defined
    by}\qquad U_a \mapsto e^B(\xi_a + i^*\theta_a)~.
\end{equation*}

\section{Gauging the WZW model with boundary}
\label{sec:wzw}

In this section we apply the preceding discussion to the gauging of a
WZW model with boundary.  We will show that we will be able to gauge
any group which can be gauged in the case without boundary, provided
that we use boundary conditions which are orbits of the group we are
trying to gauge.  Two natural classes of such gaugings are (twisted)
diagonal subgroups with boundary conditions given by (twisted)
conjugacy classes, and chiral isotropic subgroups with boundary
conditions given by cosets.  We start by setting up the notation for
computing with Lie groups.

\subsection{Some yoga about Lie groups}

Throughout this section, $G$ will denote a (connected) Lie group with
a bi-invariant metric.  We will be interested in gauging subgroups $K$
of the isometry subgroup $G \times G$.  Let $\fg$ denote the Lie
algebra and let $\left<-,-\right>$ denote the invariant inner product.
Let $\theta_L \in \Omega^1(G;\fg)$ denote the left-invariant
Maurer--Cartan form on $G$.  It is a $\fg$-valued one-form on $G$.
Identifying $T_e G \cong \fg$, we find that at the identity
$\theta_L\bigr|_e = \id$ and hence at any other point
\begin{equation*}
  \theta_L\bigr|_g = \lambda_{g^{-1}}^* \id~,
\end{equation*}
where $\lambda_g:G \to G$ denotes left-multiplication by $g$.
For matrix groups, $\theta_L\bigr|_g = g^{-1} dg$ with some abuse of
notation.  This representation is useful in computations.  For
example, one can see immediately that $\theta_L$ satisfies the
structure equation
\begin{equation*}
  d\theta_L = -\half [\theta_L, \theta_L] \in \Omega^2(G;\fg)~.
\end{equation*}
The standard three-form $H \in \Omega^3(G)$ can be written as
\begin{equation*}
  H = \tfrac16 \left< \theta_L, [\theta_L,\theta_L]\right>~.
\end{equation*}
This shows that $H$ is left-invariant.

Corresponding to every $X\in\fg$ there is a left-invariant vector
field $X^L$ obeying $X^L(e) = X$ and hence $X^L(g) = (\lambda_g)_*
X$, or equivalently
\begin{equation*}
  X^L(g) = \frac{d}{dt}\Biggr|_{t=0} \left( g e^{tX} \right)~.
\end{equation*}
It is clear that $\theta_L(X^L) = X$.  This map $X \mapsto X^L$ is
an \emph{anti}-homomorphism of Lie algebras, whence $[X^L,Y^L] = -
[X,Y]^L$, where the bracket on the left is the Lie bracket of
vector fields on $G$ and the one on the right is the one in $\fg$.

Let $j:G \to G$ denote the inverse map: $j(g) = g^{-1}$.  Define
$\theta_R = j^* \theta_L$; in other words,
\begin{equation*}
  \begin{split}
    \theta_R\bigr|_g &= j^* \theta_L\bigr|_{g^{-1}}\\
    &= j^* \lambda_g^* \id\\
    &= (\lambda_g \circ j)^* \id\\
    &= (j \circ \rho_{g^{-1}})^* \id\\
    &= \rho_{g^{-1}}^* j^* \id~,
  \end{split}
\end{equation*}
where we have used that $\lambda_g \circ j = j \circ \rho_{g^{-1}}$,
where $\rho_g:G \to G$ denotes right translation by $g$, sending $g_0
\mapsto g_0 g$.  We now notice that at the identity $j_*\bigr|_e = -
\id$, whence the same is true for $j^*$.  Therefore
\begin{equation*}
  \theta_R\bigr|_g = - \rho_{g^{-1}}^* \id~.
\end{equation*}

Similarly, let $X^R = j_* X^L$.  After an analogous calculation to the
one above we find that
\begin{equation*}
  X^R(g) = - (\rho_g)_* X = \frac{d}{dt}\Biggr|_{t=0} \left( e^{-tX}
    g\right)~.
\end{equation*}
In particular, $X^R(e) = -X$.  Notice that still $\theta_R(X^R) = X$.

Since they are related by pull-backs, $\theta_R$ obeys the same
structure equation as $\theta_L$:
\begin{equation*}
  d\theta_R = -\half [\theta_R, \theta_R]~.
\end{equation*}
In terms of $\theta_R$ we find
\begin{equation*}
  H = -\tfrac16 \left< \theta_R, [\theta_R,\theta_R]\right>~,
\end{equation*}
where the sign is due to the fact that at the identity
$\theta_R\bigr|_e = - \theta_L\bigr|_e$.  This expression shows that
$H$ is also right-invariant, hence it is bi-invariant.

Finally we collect some useful identities:
\begin{equation}
  \label{eq:RL}
  \theta_R(X^L)\bigr|_g = - \Ad_g X \qquad\text{and}\qquad
  \theta_L(X^R)\bigr|_g = - \Ad_{g^{-1}} X~,
\end{equation}
and
\begin{equation}
  \label{eq:dAd}
  d \Ad_g X = [\Ad_g X, \theta_R] \qquad\text{and}\qquad
  d \Ad_{g^{-1}} X = [\Ad_{g^{-1}} X, \theta_L]~,
\end{equation}
where $\Ad_g X = (\lambda_g \circ \rho_{g^{-1}})_* X  = g X g^{-1}$,
with some abuse of notation.

\subsection{The explicit obstructions}

Now let $K$ be a Lie group acting on $G$ by isometries, and hence
preserving $H$, since $H$ is bi-invariant.  Since the isometry group
of $G$ is $G\times G$, we have a group homomorphism $K \to G \times
G$, which in turn is uniquely characterised by two group homomorphisms
$\ell,r : K \to G$.  At the level of the Lie algebra, we also have
homomorphisms $\ell,r: \fk \to \fg$ which determine the map $\fk \to
\fg \oplus \fg$.

The action of $G\times G$ on $G$ is given by
\begin{equation*}
  (g_L,g_R) \cdot g = g_L g g_R^{-1}~.
\end{equation*}
Therefore if $X\in\fk$, and $(\ell(X),r(X))$ the corresponding element
in $\fg \oplus \fg$, then the Killing vector at $g$ is given by
\begin{equation*}
  \frac{d}{dt}\Biggr|_{t=0} \left( e^{t\ell(X)} g e^{-tr(X)} \right) =
  \frac{d}{dt}\Biggr|_{t=0} \left( e^{t\ell(X)} g \right) +
  \frac{d}{dt}\Biggr|_{t=0} \left( g e^{-tr(X)} \right) =
  - r(X)^L - \ell(X)^R~.
\end{equation*}
Let us define the Killing vector corresponding to $X$ as
\begin{equation*}
  \xi_X := r(X)^L + \ell(X)^R~,
\end{equation*}
where the sign has been changed so that $[\xi_X,\xi_Y] = \xi_{[X,Y]}$.

Let $X_a$ be a basis of $\fk$ with Killing vectors $\xi_a = r(X_a)^L +
\ell(X_a)^R$.  And let us calculate $\imath_a H$:
\begin{equation*}
  \begin{split}
    \imath_a H &= \tfrac16 \imath_{r(X_a)^L}
    \left<\theta_L,[\theta_L,\theta_L]\right> - 
    \tfrac16 \imath_{\ell(X_a)^R}
    \left<\theta_R,[\theta_R,\theta_R]\right> \\
    &= \half \left<r(X_a), [\theta_L,\theta_L]\right> - 
    \half \left<\ell(X_a), [\theta_R,\theta_R]\right>\\
    &= \left<\ell(X_a), d\theta_R\right> - 
    \left<r(X_a), d\theta_L\right>\\
    &= d \theta_a~,
  \end{split}
\end{equation*}
where
\begin{equation*}
  \theta_a = \left<\ell(X_a), \theta_R\right> - \left<r(X_a),
    \theta_L\right>~.
\end{equation*}

We next check that $\eL_a \theta_b = f_{ab}{}^c \theta_c$, for which
we first compute
\begin{equation*}
  \imath_a \theta_b = \left<\ell(X_a),\ell(X_b)\right> - 
  \left<\ell(X_b), \Ad_g r(X_a)\right> - \left<r(X_a),r(X_b)\right> +
  \left<r(X_b), \Ad_{g^{-1}} \ell(X_a)\right>~,
\end{equation*}
where we have used equations \eqref{eq:RL}, and
\begin{multline*}
  \imath_a\imath_b H = - \left<\ell(X_b), [\ell(X_a),\theta_R]\right>
  + \left<\ell(X_b), [\Ad_g r(X_a),\theta_R]\right>\\
  + \left<r(X_b), [r(X_a),\theta_L]\right> -
  \left<r(X_b), [\Ad_{g^{-1}} \ell(X_a),\theta_L]\right>~.
\end{multline*}

Continuing with the calculation, we find, using \eqref{eq:dAd}, that
\begin{equation*}
  \begin{split}
    \eL_a \theta_b &= d \imath_a \theta_b + \imath_a \imath_b H\\
    &= - \left<\ell(X_b), d \Ad_g r(X_a)\right> + \left<r(X_b), d
      \Ad_{g^{-1}} \ell(X_a)\right>
    + \left<[\ell(X_a), \ell(X_b)], \theta_R\right>\\
    & \qquad + \left<\ell(X_b), [\Ad_g r(X_a),\theta_R]\right> -
    \left<[r(X_a), r(X_b)], \theta_L\right> -
    \left<r(X_b), [\Ad_{g^{-1}} \ell(X_a),\theta_L]\right>\\
    &= - \left<\ell(X_b), [\Ad_g r(X_a),\theta_R] \right> +
    \left<r(X_b), [\Ad_{g^{-1}} \ell(X_a), \theta_L] \right>
    + \left<[\ell(X_a), \ell(X_b)], \theta_R\right>\\
    & \qquad + \left<\ell(X_b), [\Ad_g r(X_a),\theta_R]\right> -
    \left<[r(X_a), r(X_b)], \theta_L\right> -
    \left<r(X_b), [\Ad_{g^{-1}} \ell(X_a),\theta_L]\right>\\
    &= \left<[\ell(X_a), \ell(X_b)], \theta_R\right> - \left<[r(X_a),
      r(X_b)], \theta_L\right>~,
  \end{split}
\end{equation*}
which yields the desired result after using that $\ell$ and $r$ are
Lie algebra homomorphisms.

Finally the obstruction $\imath_a\theta_b + \imath_b \theta_a = 0$
translates into
\begin{equation}
  \label{eq:obsWZW}
  \left<\ell(X_a),\ell(X_b)\right> = \left<r(X_a),r(X_b)\right>~.  
\end{equation}
In other words, the scalar products induced on $\fk$ via $\ell,r:\fk
\to \fg$ should balance.

There are (at least) two natural ways in which this can be achieved:
one is by a diagonal gauging in which $\ell=r$, or more generally a
twisted diagonal gauging in which $\ell = r \circ \tau$, for some
orthogonal automorphism $\tau$ of $\fg$.  The other general class are
chiral gaugings where, say, $r=0$ and where $\ell(\fk) \subset \fg$ is
an isotropic subalgebra.  We will discuss both cases at the end of
this section, once we have shown that any group for which the WZW
model without boundary can be gauged, can also be gauged when the
boundary is present, provided that we choose appropriately symmetric
boundary conditions.

\subsection{The boundary offers no new obstructions}

Let us consider gauging a group $K$ acting via isometries on $G$.  As
discussed above this action is defined by group homomorphisms
$\ell,r:K \to G$.  The action of $k \in K$ on $G$ is denoted
\begin{equation*}
  \Ad_k^{\ell,r} : G \to G \qquad\text{where}\qquad
  \Ad_k^{\ell,r} (g) = \ell(k) g r(k)^{-1}~.
\end{equation*}
The possible boundary conditions are the orbits of this action:
\begin{equation*}
  C^{\ell,r}(g_0) = \left\{ \ell(k) g_0 r(k)^{-1} \mid k \in K
  \right\}~.
\end{equation*}
Fix $g_0\in G$ and let $Y:=C^{\ell,r}(g_0)$. We will denote $i:Y \to
G$ the canonical embedding.

We will first show that $i^*H = dB$ for some two-form $B\in
\Omega^2(Y)$.  To show this and for later usage, it is convenient to
write down the pull-backs of the Maurer--Cartan forms to $Y$.  Letting
$g = \Ad^{\ell,r}_k g_0$ and letting $\vartheta = dkk^{-1}$, we find
\begin{equation}
  \label{eq:PBMC}
  i^*\theta_L\bigr|_g = (\Ad_{g^{-1}} \ell - r) \vartheta
  \qquad\text{and}\qquad
  i^*\theta_R\bigr|_g = (\Ad_g r - \ell) \vartheta~,
\end{equation}
where we let $\ell,r$ also denote the Lie algebra homomorphisms $\fk
\to \fg$.

A short calculation, using equation~\eqref{eq:obsWZW}, shows that
\begin{equation*}
  i^* H\bigr|_g  = \left< \alpha \vartheta, rd\vartheta\right> -
  \left< r\vartheta, \alpha d\vartheta\right>~,
\end{equation*}
where we have introduced the notation $\alpha = \Ad_{g^{-1}} \ell$.
Now consider
\begin{equation*}
  B = - \left< r \vartheta, \alpha \vartheta\right> = - \left< r \vartheta,
    i^*\theta_L\right>~.
\end{equation*}
Differentiating, one gets
\begin{equation*}
  \begin{split}
    dB &= - \left< r d\vartheta, i^*\theta_L\right> +
    \left< r \vartheta, i^*d\theta_L\right>\\
    &= - \left< r d\vartheta, i^*\theta_L\right> -
    \half \left<\vartheta, [i^*\theta_L,i^*\theta_L]\right>~.
  \end{split}
\end{equation*}
We now use the first equation in \eqref{eq:PBMC}, expand and use
that $r$ and $\alpha$ are Lie algebra homomorphisms, to obtain that
\begin{equation*}
  dB = \left<r d\vartheta, \alpha \vartheta\right> - \left<r \vartheta, \alpha
    d\vartheta\right> = i^* H~.
\end{equation*}

We now show that $B$ is $K$-invariant.  Indeed, if $x\in K$, then
$\Ad^{\ell,r}_x g = \ell(xk) g_0 r(xk)^{-1}$, whence $k$ changes by
left multiplication by $x$, so that in turn $\vartheta$ changes to
$\Ad_x \vartheta$.  Finally we calculate
\begin{equation*}
  \begin{split}
    \left(\Ad_x^{\ell,r}\right)^* B_{\Ad^{\ell,r}_x g} & = - \left<
      r \left(\Ad_x^{\ell,r}\right)^* \vartheta, 
        \Ad_{(\ell(x)gr(x)^{-1})^{-1}} \ell \left(\Ad_x^{\ell,r}\right)^*
        \vartheta\right>\\
    & = - \left< r \Ad_x \vartheta, \Ad_{r(x)} \Ad_{g^{-1}}
        \Ad_{\ell(x)^{-1}} \ell \Ad_x \vartheta\right>\\
    &= - \left< \Ad_{r(x)^{-1}} r \Ad_x \vartheta,  \Ad_{g^{-1}}
      \Ad_{\ell(x)^{-1}} \ell  \Ad_x \vartheta\right>\\
    &= B_{g}~,
  \end{split}
\end{equation*}
where we have used that the metric is $\Ad$-invariant and that for any
homomorphism $\tau$, whence in particular for $\ell$ and $r$,
\begin{equation*}
  \Ad_{\tau(x)^{-1}} \tau  \Ad_x  = \tau~,
\end{equation*}
where the $\Ad$ in the left is in $K$ and the one in the right is in
$G$.  This remark also applies to the above calculation, where we hope
no confusion should result as a consequence of this ambiguity in the
notation.

This means that the relative WZ term with these boundary
conditions is invariant under the (global) action of $K$ and we can
address the problem of gauging it.  We will prove that we can do so.

Given $X_a \in \fk$, the corresponding Killing vector is
\begin{equation*}
  \xi_a = \ell(X_a)^R + r(X_a)^L~,
\end{equation*}
whence $\imath_a H = d \theta_a$, where
\begin{equation*}
  \theta_a = \left<\ell(X_a), \theta_R\right> - \left<r(X_a),
    \theta_L\right>~.
\end{equation*}

Pulling back $\theta_a$ to $Y$ we find
\begin{equation*}
  \begin{split}
    i^* \theta_a\bigr|_g &= \left<\ell(X_a), i^*\theta_R\right> -
    \left<r(X_a), i^*\theta_L\right>\\
    &= \left<\ell(X_a), (\Ad_g r - \ell) \vartheta\right> - \left<r(X_a),
      (\Ad_{g^{-1}} \ell - r) \vartheta\right>\\
    &= \left<\alpha(X_a), r \vartheta\right> - \left<r(X_a),
      \alpha \vartheta\right>~,
  \end{split}
\end{equation*}
where we have again used equation~\eqref{eq:obsWZW}.

On the other hand, in the chosen parametrisation for $g = \Ad^{\ell,r}_k
g_0$, $\xi_a$ generates
\begin{equation*}
  k \mapsto e^{-t X_a} k
\end{equation*}
whence it corresponds to $X_a^R$.  Since $\vartheta$ pulls back to
$-\theta_R$ on $K$, we see that $\imath_a \vartheta = - X_a$.  Using
this we find,
\begin{equation*}
  \begin{split}
  \imath_a B &= - \left< r \imath_a\vartheta, \alpha
    \vartheta\right>  + \left<r \vartheta, \alpha
    \imath_a\vartheta\right>\\
  &= \left<r(X_a), \alpha\vartheta\right> - \left<\alpha X_a,
    r \vartheta\right>\\
  & = - i^*\theta_a~.
  \end{split}
\end{equation*}

From equation~\eqref{eq:2drelative}, we see that $dh_a = 0$, whence
$h_a$ defines an element $h \in \fk^*$.  The equivariance condition
$\eL_a h_b = f_{ab}{}^c h_c$ means that $h$ lies in the annihilator
$[\fk,\fk]^o$ of the first derived ideal.

Since $dh_a=0$, the boundary Lie algebroid in this case is the
canonical Lie algebroid associated to the action of $\fg$ on $Y$.

In summary, we have proved that any subgroup which can be gauged in
the WZW model without boundary, can still be gauged in the presence of
a boundary provided that the boundary conditions are chosen
appropriately, namely they are invariant under the group we are trying
to gauge, or more geometrically, they consist of orbits of the action.

It is now a simple matter to write down the gauged WZ term by
specialising equation~\eqref{eq:GRWZ} to the present case.  Doing so
we obtain
\begin{multline}
  \label{eq:GRWZW}
  S_{\text{grWZW}} = \int_{\widetilde\Sigma} \tfrac13 \Tr (g^{-1}
  dg)^3 + \int_\Sigma \Tr \left( g^{-1}\ell(A_z) g r(A_{\bar z}) -
    g^{-1} \ell(A_{\bar z}) g r(A_z)
  \right) d^2z\\
  - \int_\Sigma \Tr \left( \ell(A_z) \partial_{\bar z}g g^{-1} -
    \ell(A_{\bar z}) \partial_z g g^{-1} + r(A_z) g^{-1}
    \partial_{\bar z}g - r(A_{\bar z})
    g^{-1} \partial_z g \right) d^2z\\
  - \int_\Delta \Tr \left( g_0 r(k)^{-1} \partial_\zeta r(k) g_0^{-1}
    \ell(k)^{-1} \partial_{\bar\zeta}\ell(k) - g_0 r(k)^{-1}
    \partial_{\bar\zeta} r(k) g_0^{-1} \ell(k)^{-1}
    \partial_\zeta\ell(k)\right) d^2\zeta\\
  + \int_{\partial\Sigma} h(A_x) dx~,
\end{multline}
where again $\widetilde\Sigma$ is a three-manifold with boundary
$\partial\widetilde\Sigma = \Sigma + \Delta$, where $z$ is a local
complex coordinate on $\Sigma$, $\zeta$ a local complex coordinate on
$\Delta$ and $x$ is a local real coordinate on $\partial\Sigma$.  In
this formula we have used $\Tr$ to mean the invariant scalar product
in the Lie algebra.  This inner product encodes the information on the
``level'' of the WZW model.  The symbol $g$ denotes both the map
$g:\Sigma \to G$ as well as its extension to $\widetilde\Sigma$,
whereas in $\Delta$, we parametrise $g = \ell(k)g_0 r(k)^{-1}$ in
terms of a map $k:\Delta \to K$.  The last two terms in
equation~\eqref{eq:GRWZW} are due to the presence of the boundary.

\subsection{Two examples}

Now we consider the two natural examples of gaugings.

\subsubsection{(Twisted) Diagonal gaugings}

First we have $G$ embedded diagonally in $G\times G$: $\ell=r=\id$ or
more generally as the graph of an orthogonal automorphism $\tau: G\to
G$.  The possible boundary conditions are the orbits of this action
which, in the general case of the graph of an automorphism are twisted
conjugacy classes
\begin{equation*}
  C^\tau(g_0) = \left\{ \tau(g) g_0 g^{-1} \mid g \in G \right\}~.
\end{equation*}
Such orbits are well-known in the context of D-branes on Lie groups
\cite{AS,FFFS,SDnotes}.

This corresponds to $\ell = \tau$ and $r=\id$ in the previous
section and the results from that section can be immediately
imported.  We have $i^*H = dB$, where \cite{q0} (see also
\cite{AS,Gaw} for the untwisted case)
\begin{equation*}
  B = - \left< \vartheta, \alpha \vartheta\right>~,
\end{equation*}
where now $\alpha = \Ad_{g^{-1}} \tau$.  As we showed above, $B$ is
invariant under the twisted adjoint action, meaning that the relative
WZ term with such boundary conditions is invariant under the
global action of $G$ via twisted conjugation.  As shown above in a
more general context, this global action can be gauged.

The action can be read off from equation~\eqref{eq:GRWZW} by putting
$r=\id$ and $\ell = \tau$.  The new terms due to the boundary are now
\begin{multline*}
  S_{\text{grWZW}} = \cdots
  - \int_\Delta \Tr \left( g_0 k^{-1} \partial_\zeta k g_0^{-1}
    \tau(k)^{-1} \partial_{\bar\zeta}\tau(k) - g_0 k^{-1}
    \partial_{\bar\zeta} k g_0^{-1} \tau(k)^{-1}
    \partial_\zeta\tau(k)\right) d^2\zeta\\
  + \int_{\partial\Sigma} h(A_x) dx~,
\end{multline*}
The boundary piece in the gauged action depends on an element $h \in
[\fg,\fg]^o$.  Using the invariant scalar product, $h$ defines an
element in the centre of $\fg$.  In other words, the gauge field
$h(A_x)$ on the boundary takes values in the centre of the Lie
algebra, in particular it is always abelian.  Semisimple Lie algebras
have no centre, which explains why this term in absent in
\cite{GawedzkiCoset,ElitzurSarkissian,KRWZ}.

\subsubsection{Chiral isotropic gaugings}

Another set of subgroups which can be gauged are chiral subgroups where,
say, $r =0$.  In this case, the condition to be able
to gauge the WZ term (without boundary) is that
\begin{equation*}
  \left<\ell(X_a),\ell(X_b)\right> = 0~.
\end{equation*}
One class of such subgroups is the following.  Let $\fg$ be a
maximally noncompact real form of a complex simple Lie algebra and let
\begin{equation*}
  \fg = \fh \oplus \fn_+ \oplus \fn_-
\end{equation*}
be a maximal toral decomposition, where $\fh$ is the Cartan subalgebra
and
\begin{equation*}
  \fn_\pm = \bigoplus_{\alpha \in \Phi_\pm} \fg_\alpha~,
\end{equation*}
where $\Phi = \Phi_+ \sqcup \Phi_-$ is the root system, with
$\Phi_+$ (resp. $\Phi_-$) the positive (resp. negative) roots, and
$\fg_\alpha$ the one-dimensional root space with root $\alpha$.  For
example, we can take $\fg = \fsl(n,\RR)$ to be the Lie algebra of
traceless $n\times n$ matrices with real entries, $\fh$ the subalgebra
of diagonal matrices, and $\fn_\pm$ the subalgebras of strictly
triangular matrices.  We let $N_\pm \subset G$ denote the subgroup
with Lie algebra $\fn_\pm$.

The subalgebra $\fn_+$ is isotropic under the Killing form and we
can try to gauge the corresponding group as a chiral gauging: $\fn_+
\to \fg \oplus \fg$ sending $X \to (X,0)$, for example.  The possible
boundary conditions are orbits of $N_+$ under this action, which here
correspond to left cosets
\begin{equation*}
  Y := N_+ g_0 = \left\{ h g_0 \mid h \in N_+ \right\}~.
\end{equation*}
Let $i: Y \to G$ denote the embedding.  As we saw above, we can gauge
such a symmetry with these boundary conditions.  It is perhaps
instructive to redo the calculation in this case, since as we will see
the two-form $B$ can be taken to be zero.  As a result the boundary
WZW model is unchanged and the gauged model only receives a
contribution from the boundary of the world-sheet.

Let $g = h g_0$ with $h\in N_+$.  Then we find
\begin{equation*}
  i^* \theta_L\bigr|_g = \Ad_{g^{-1}} \vartheta
  \qquad\text{and}\qquad
  i^* \theta_R\bigr|_g = - \vartheta~,
\end{equation*}
where $\vartheta = dh h^{-1}$ as before.  Then
\begin{equation*}
  i^*H = \half \left<\vartheta, [\vartheta, \vartheta]\right> = 0~,
\end{equation*}
since $\vartheta$ is $\fn_+$-valued, and this subalgebra is
isotropic.  This means that we can take $B = 0$.

If $X_a$ is a basis for $\fn_+$, the corresponding Killing vector is
$X_a^R$.  Then
\begin{equation*}
  \begin{split}
    \imath_a H &= - \half \left< X_a, [\theta_R, \theta_R]\right>\\
    &= \left< X_a, d\theta_R\right>\\
    & = d \left<X_a, \theta_R\right>~,
  \end{split}
\end{equation*}
whence $\theta_a = \left<X_a, \theta_R\right>$.  Pulling back to $Y$,
we find
\begin{equation*}
  i^* \theta_a = \left<X_a, i^*\theta_R\right> = - \left<X_a,
    \vartheta\right> = 0
\end{equation*}
again by isotropy of $\fn_+$.

Therefore $\imath_aB + i^*\theta_a =0$ identically and we again
have that $dh_a=0$, whence it defines an element $h \in \fn_+^*$ which
annihilates $[\fn_+,\fn_+]$.  In other words, it necessarily
annihilates all the non-simple roots.  Hence the gauge field on the
boundary is a linear combination $h_i A^i$ where $i$ runs through the
simple roots.

The gauged WZ term can be read off from equation~\eqref{eq:GRWZW} and
we obtain
\begin{equation*}
  S_{\text{grWZW}} = \int_{\widetilde\Sigma} \tfrac13 \Tr (g^{-1}
  dg)^3 - \int_\Sigma \Tr \left( \ell(A_z) \partial_{\bar z}g g^{-1} -
    \ell(A_{\bar z}) \partial_z g g^{-1} \right) d^2z +
  \int_{\partial\Sigma} h(A_x) dx~.
\end{equation*}

These chiral gaugings are well-known in the context of integrable
models as they provide a WZW realisation of the Drinfel'd--Sokolov
reduction.  The gauged WZW models with boundary could play a role in
the construction of boundary integrable models and we hope to return
to this topic elsewhere.

\section*{Acknowledgments}

To some extent this work is a natural continuation of both
\cite{FSsigma,FSec} and \cite{FSrc}, which JMF wrote together with
Sonia Stanciu.  JMF was reminded of this programme in a conversation
with Takashi Kimura at the IAS in the Spring of 2003 and it is his
pleasure to thank him.  The project received new life when the authors
met briefly in Edinburgh and then in DAMTP (Cambridge) in the Spring
and Summer of 2003.  NM would like to thank the LMS for funding his
visit to Edinburgh in the context of the North British Mathematical
Physics Seminar and Hugh Osborn for the invitation to visit Cambridge.
After a lengthy hiatus, this work was given its final push in the
Spring of 2005 during a visit of JMF to the Laboratoire de
Mathématiques et Physique Théorique de l'Université de Tours, whom he
would like to thank for hospitality and support.  The work was
finished during a visit of JMF to the IHÉS, whom he would like to
thank for hospitality and support.  JMF would also like to thank Chris
Hull for reminding him about the relation between Courant brackets and
sigma models.  We welcome the opportunity to thank an anonymous
referee for some useful comments.

Last, but certainly not least, JMF would like to thank da tutto il
cuore l'Ospedale Merzagora--Petrini per Bambini Anziani for their
care, hospitality and support beyond the call of duty during the final
stages of this work.

\appendix

\section{$G$-equivariant relative cohomology}
\label{app:equivrel}

Let $i:Y \to X$ be a submanifold and let $G$ be a connected Lie group
acting on $X$ preserving $Y$, so that $i$ is $G$-equivariant.  With
this data we can define at least two notions of ``$G$-equivariant
cohomology of $X$ relative to $Y$'' depending on which model we choose
for the relative de Rham complex.  In the absence of further
assumptions --- e.g., compactness of the group --- both choices give
different theories, and only one of them is relevant to the physical
problem at hand.  In this Appendix we try to explain this situation.

Recall that there are two complexes computing the de Rham cohomology
of $X$ relative to $Y$.  Let $i^*: \Omega(X) \to \Omega(Y)$ denote the
operation of pulling back forms from $X$ to $Y$ and let $\Omega(X,Y)$
denote its kernel.  We thus have a short exact sequence of complexes
\begin{equation}
  \label{eq:RdR1}
    \begin{CD}
    0 @>>> \Omega^\bullet(X,Y) @>>> \Omega^\bullet(X) @>i^*>>
    \Omega^\bullet(Y) @>>> 0~.
  \end{CD}
\end{equation}
Exactness at the right is simply the possibility to extend a form on
$Y$ smoothly to all of $X$.  This short exact sequence gives a long
exact sequence in cohomology whose coboundary map $H^p(Y) \to
H^{p+1}(X,Y)$ is obtained by sending $[\omega]\in H^p(Y)$ to
$[d\widetilde \omega]$, where $\widetilde \omega$ is an extension of
$\omega$ to $X$.  The cohomology space $H^*(Y,X)$ is the de Rham
cohomology of $X$ relative to $Y$.

One can relax the notion that a relative form should vanish when
pulled back to $Y$ to the level of cohomology alone; that is, that a
relative cocycle need not vanish when pulled back to $Y$ but that it
should be exact there.  This suggests defining a different complex
\begin{equation*}
  \widetilde\Omega^p(X,Y) := \Omega^p(X) \oplus \Omega^{p-1}(Y)
\end{equation*}
with differential $\widetilde d : \widetilde\Omega^p(X,Y) \to
\widetilde\Omega^{p+1}(X,Y)$ defined by
\begin{equation*}
  \widetilde d (\omega, \theta) = (d\omega, d\theta + (-1)^p
  i^*\omega)~.
\end{equation*}
A cocycle is then a closed form $\omega$ whose pull-back to $Y$ is
exact.

This complex also fits in a short exact sequence
\begin{equation}
  \label{eq:RdR2}
    \begin{CD}
    0 @>>> \Omega^{\bullet-1}(Y) @>>> \widetilde \Omega^\bullet(X,Y)
    @>>> \Omega^\bullet(X) @>>> 0~,
  \end{CD}
\end{equation}
where the first map is simply the inclusion into the second factor
$\theta \mapsto (0,\theta)$ and the second map is the projection onto
the first $(\omega,\theta) \mapsto \omega$.  It is possible to match
the resulting long exact sequence in cohomology with the one coming
from the first complex and in this way show that the map $\Omega(X,Y)
\to \widetilde\Omega(X,Y)$ defined by $\omega \mapsto (\omega, 0)$ is
a quasi-isomorphism, whence the two complexes compute the same
cohomology.

Now suppose that $G$ acts on $X$ and $Y$ with $i:Y \to X$
$G$-equivariant.  Then both relative de Rham complexes become
$G$-dgas in natural ways.  First of all, since $G$ acts on $Y$, the
Killing vectors are tangent to $Y$ when restricted to $Y$, whence the
derivations $\imath_a$ and $\eL_a$ restrict to $\Omega(X,Y)$.
Similarly for the second relative de Rham complex
$\widetilde\Omega(X,Y)$, we define $\imath_a(\omega,\theta) =
(\imath_a \omega,\imath_a\theta)$ and $\eL_a (\omega,\theta) =
(\eL_a\omega,\eL_a\theta)$.  One checks that again $\eL_a =
\widetilde d \imath_a + \imath_a \widetilde d$.  Moreover the
quasi-isomorphism $\Omega(X,Y) \to \widetilde\Omega(X,Y)$ is
a morphism of $G$-dgas.  This means, for example, that we get a map
between the cohomologies  of the invariant complexes $H(X,Y)^G \to
\widetilde H(X,Y)^G$, which however need not be an isomorphism.
Indeed, trying to prove the isomorphism as for the case of the
relative de Rham complexes, we come across the fact that whereas
we still have a short exact sequence of invariant complexes
\begin{equation}
  \label{eq:RdR2G}
    \begin{CD}
    0 @>>> \Omega^{\bullet-1}(Y)^G @>>> \widetilde \Omega^\bullet(X,Y)^G
    @>>> \Omega^\bullet(X)^G @>>> 0~,
  \end{CD}
\end{equation}
the other sequence
\begin{equation}
  \label{eq:RdR1G}
    \begin{CD}
    0 @>>> \Omega^\bullet(X,Y)^G @>>> \Omega^\bullet(X)^G @>i^*>>
    \Omega^\bullet(Y)^G
  \end{CD}
\end{equation}
now fails to be right-exact in general, because an invariant form on
$Y$ need not extend to an \emph{invariant} form on all of $X$.  This
will be the case if $G$ is compact, by averaging, for example, but is
not the case for general $G$.

This means that we have two possible definitions of $G$-equivariant
cohomology of $X$ relative to $Y$, depending on which complex we take
for relative cohomology.  On the one hand we have
\begin{equation*}
  \Omega_G(X,Y):= \left(W(\fg) \otimes \Omega(X,Y)\right)_{\text{basic}}
\end{equation*}
which is a subcomplex of $\Omega_G(X)$.  On the other hand, we have
\begin{equation*}
  \widetilde\Omega_G(X,Y):= \left(W(\fg) \otimes
    \widetilde\Omega(X,Y)\right)_{\text{basic}}~.
\end{equation*}
In general both complexes compute different cohomologies; although it
is clear that it is this latter complex which is relevant for our
problem.

One might worry about another source of ambiguity coming from the
order of operations; that is, whether we first ``relativise'' or
``equivariantise'' the complex.  Starting with the $G$-dga
$\Omega(X)$, one can first consider its equivariant cohomology and
then make this relative to $Y$, or one can first consider the relative
cohomology and then make this equivariant.  It is not hard to see, 
however, that for either of the two models of the relative de Rham
complex considered above, both procedures yield isomorphic complexes.

Having identified the right equivariant relative de Rham complex, it
should be possible, as in \cite{FSsigma,FSec}, to derive a priori
vanishing theorems which guarantee the absence of obstructions to
gauging for certain types of groups and certain geometries.

\bibliographystyle{utphys}
\bibliography{AdS3,Geometry}

\end{document}